\documentclass[%
reprint,
superscriptaddress,
amsmath,amssymb,
aps,
pra,
]{revtex4-2}

\usepackage[
CJKbookmarks=false,
colorlinks,
breaklinks=true,
linkcolor=blue,
anchorcolor=blue,
citecolor=blue,
urlcolor=blue,
]{hyperref}

\usepackage[normalem]{ulem} 

\usepackage{bibunits}
\usepackage{graphicx}
\usepackage{dcolumn}
\usepackage{bm}

\begin{document}


\title[]{Implementing rapid reset of superconducting qubits with on-chip dissipation}

\newcommand{\BAQIS}{\affiliation{1}{Beijing Key Laboratory of Fault-Tolerant Quantum Computing, \\
Beijing Academy of Quantum Information Sciences, Beijing 100193, China}}
\newcommand{\IOP}{\affiliation{2}{Beijing National Laboratory for Condensed Matter Physics,\\
Institute of Physics, Chinese Academy of Sciences, Beijing 100190, China}}
\newcommand{\UCAS}{\affiliation{3}{University of Chinese Academy of Sciences, Beijing 101408, China}}

\author{Feiyu Li}
\thanks{ These authors contributed equally to this work.} 
\affiliation{\BAQIS}
\affiliation{\IOP}
\affiliation{\UCAS}

\author{Yang Liu}
\thanks{ These authors contributed equally to this work.}
\affiliation{\BAQIS}

\author{Zhen Yang}
\thanks{ These authors contributed equally to this work.}
\affiliation{\BAQIS}

\author{Zhenxing Liu}
\affiliation{\BAQIS}

\author{Jiayu Ding}
\affiliation{\BAQIS}

\author{Yang Gao}
\affiliation{\BAQIS}
\affiliation{\IOP}
\affiliation{\UCAS}

\author{Xuehui Liang}
\affiliation{\BAQIS}

\author{Weijie Sun}
\affiliation{\BAQIS}

\author{Yanjun Ma}
\affiliation{\BAQIS}

\author{Chuhong Yang}
\affiliation{\BAQIS}

\author{Guangming Xue}
\email{xuegm@baqis.ac.cn}
\affiliation{\BAQIS}

\author{He Wang}
\email{wanghe@baqis.ac.cn}
\affiliation{\BAQIS}

\author{Fei Yan}
\email{yanfei@baqis.ac.cn}
\affiliation{\BAQIS}

\date{\today}

\begin{abstract}
Rapid and high-fidelity qubit reset is essential for quantum state initialization and quantum error correction. Conventional reset protocols for superconducting qubits often rely on dissipative elements connected to external circuitry, exposing qubits to thermal noise from higher-temperature environments. Here, we experimentally demonstrate a scalable approach to qubit reset and leakage removal over multiple transmon qubits using dedicated dissipation units comprising resonators coupled to on-chip resistive loads. This architecture provides Purcell protection while isolating the qubits from warmer external circuitry. We achieve a median reset time constant of approximately 48~ns for the $|g\rangle-|e\rangle$ transition, with an effective bath temperature of 30 mK. A 400-ns reset pulse yields residual excited-state populations of $0.29\%$ and $0.67\%$ for qubits initialized in $|g\rangle$ and $|e\rangle$, respectively. We further demonstrate efficient leakage removal from the $|f\rangle$ state, reducing its residual population to $0.51\%$. These results establish a simple, reproducible, and easy-to-integrate approach to dissipation engineering in superconducting quantum processors.
\end{abstract}


\maketitle

Initializing qubits into a well-defined ground state is a fundamental requirement for quantum computation~\cite{divincenzo2000physical}.
In superconducting quantum processors, residual excitation and leakage can persist across repeated error-correction cycles and compromise logical performance, motivating rapid reset and leakage removal~\cite{mcewen2021removing,battistel2021hardware,miao2023overcoming,marques2023all,yang2024coupler,lacroix2025fast}.
Implementing these protocols across multi-qubit processors has helped enable the suppression of logical error rates with increasing code distance in recent surface-code experiments~\cite{google2025quantum,he2025experimental}.
In circuit quantum electrodynamics~\cite{blais2021circuit}, qubit reset can be implemented through engineered dissipation, with dissipative terminations located either off chip or on chip.
Approaches using external dissipative terminations typically employ a dissipative resonator shared between qubit readout and reset~\cite{reed2010fast,geerlings2013demonstrating,egger2018pulsed,magnard2018fast,zhou2021rapid,mcewen2021removing,battistel2021hardware,miao2023overcoming,marques2023all,lacroix2025fast,gu2026multimode}, or a dedicated dissipative mode~\cite{ding2025multipurpose,zhao2026broadband}, both coupled to external microwave circuitry.
However, these approaches can increase complexity in design and expose qubits to thermal noise propagating from higher-temperature stages, potentially raising their effective bath temperature.
Alternatively, on-chip approaches employ local dissipative elements, such as quantum circuit refrigerators~\cite{yoshioka2021fast,sevriuk2022initial,yoshioka2023active} and resistive loads~\cite{tuorila2017efficient,partanen2018flux}.
Local resistive termination can eliminate the need for dedicated cryogenic ports and wiring, offering a promising route toward scalable implementation. Yet, experimental demonstrations combining rapid reset with such an integration-friendly architecture remain limited.

\begin{figure*}[!t]
\centering
\includegraphics[width=\textwidth]{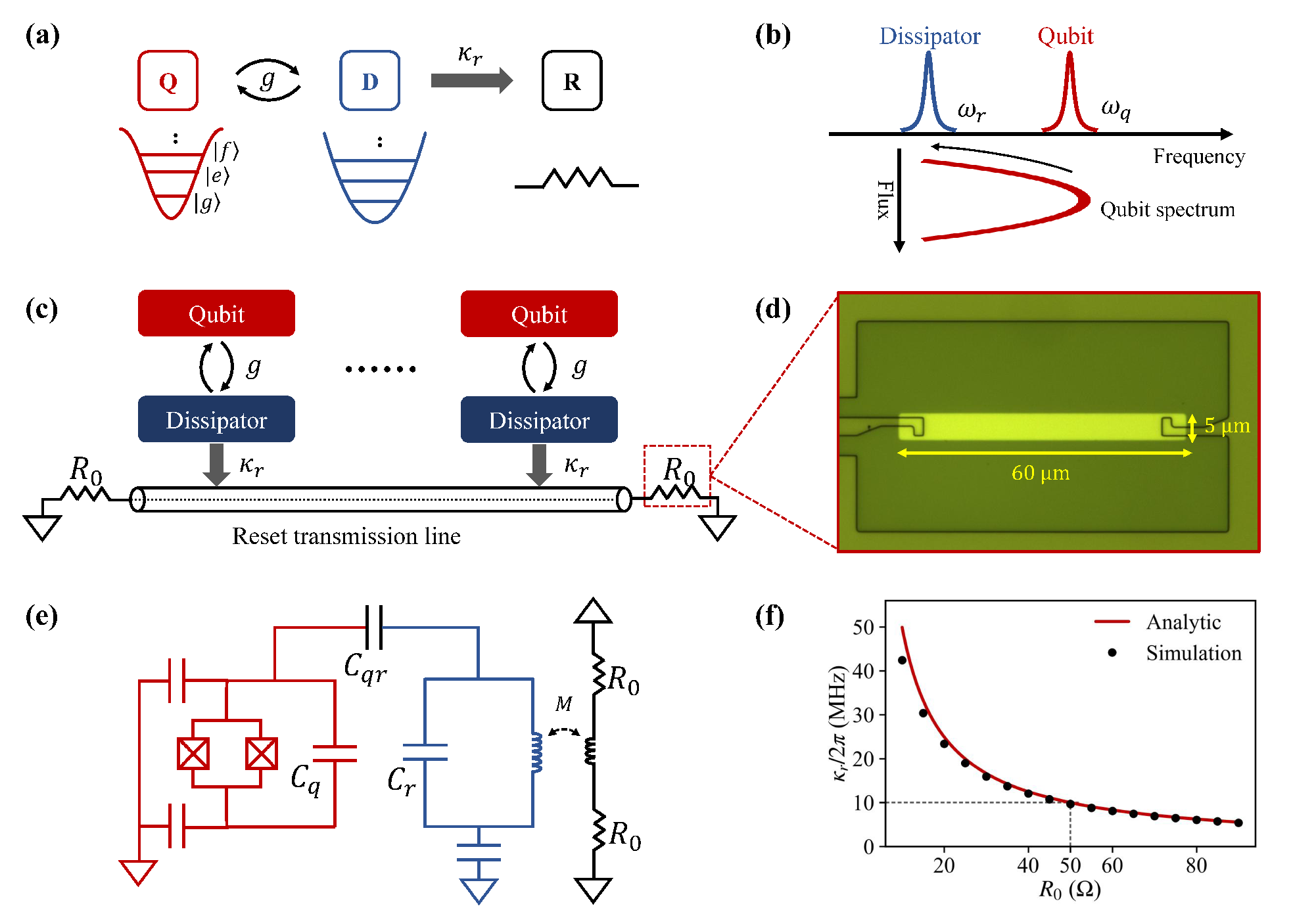}
\caption{On-chip dissipation architecture for qubit reset.
(a)~A qubit (Q) couples with strength $g$ to a dissipative resonator (D), which loses energy to an on-chip resistive load (R) at rate $\kappa_r$.
(b)~Frequency arrangement and reset mechanism. Flux tuning brings the qubit frequency $\omega_q$ into resonance with the dissipator frequency $\omega_r$ to activate reset.
(c)~Schematic of the shared reset architecture. Each qubit is capacitively coupled to a dedicated dissipator; six dissipators share a reset transmission line terminated at both ends by nominally matched resistive loads $R_0$.
(d)~Optical micrograph of a $60\times 5~\mu\mathrm{m}$ palladium--gold (PdAu) resistive termination.
(e)~Lumped-element equivalent circuit. The qubit and dissipator are coupled through $C_{qr}$, and the dissipator couples inductively to the load circuit through mutual inductance $M$.
(f)~$\kappa_r/2\pi$ versus $R_0$ from the analytical model (red curve) and finite-element electromagnetic simulations (black circles), for $\omega_r/2\pi=3.7~\mathrm{GHz}$ and $M=140~\mathrm{pH}$.
Dashed lines mark the matched design point, $R_0=50~\Omega$ and $\kappa_r/2\pi=10.0~\mathrm{MHz}$.}
\label{fig:concept}
\end{figure*}

Here, we experimentally demonstrate a qubit reset protocol based on an on-chip dissipation architecture. The underlying concept is simple: each qubit is coupled to a dedicated dissipative resonator (hereafter referred to as a dissipator), which provides a pathway for qubit excitations to be absorbed by an on-chip resistive load, as illustrated in Fig.~\ref{fig:concept}(a). The qubit frequency $\omega_q$, dissipator frequency $\omega_r$, coupling strength $g$, and dissipator decay rate $\kappa_r$ are designed to enable rapid qubit relaxation when resonant with the dissipator while suppressing Purcell decay during off-resonant coherent operations. By placing the dissipator frequency below the optimal operating frequencies of a tunable transmon qubit, rapid reset can be activated by applying a flux pulse that tunes the qubit into resonance with the dissipator [Fig.~\ref{fig:concept}(b)]. The dissipator operates passively and requires no additional control, simplifying the hardware cost.

In our device [Fig.~\ref{fig:concept}(c)], each transmon qubit is capacitively coupled to a dedicated $\lambda/4$ transmission-line resonator, with both fabricated from tantalum. Multiple dissipators are coupled to a shared reset transmission line terminated at both ends by impedance-matched palladium–gold (PdAu) thin-film resistors~\cite{yohannes2023high,olaya2023nb}.
The transmission line allows the lossy normal-metal terminations to be spatially separated from the qubits and other coherent circuit elements. 
The resistive terminations consist of a $45$-nm-thick PdAu alloy film (the light part in Fig.~\ref{fig:concept}(d)) with a Pd:Au mass ratio of $1:1$, deposited by thermal evaporation after ion milling to remove the native oxide from the tantalum surface. Figure~\ref{fig:concept}(e) presents the corresponding lumped-element equivalent circuit, in which the dissipator is inductively coupled to the reset transmission line through a mutual inductance.

In the lumped-element model, mutual coupling $M$ to the two terminations introduces a reflected series resistance $R_s=(\omega_r M)^2/R_{\rm eff}$ in the dissipator, where $R_{\rm eff}=2R_0$ is the effective resistance of the load circuit.
The equivalent dissipator inductance referred to the short-end current is $L=\pi Z_r/(4\omega_r)$, where $Z_r=50~\Omega$ is the resonator characteristic impedance. The decay rate $\kappa_r=R_s/L$ is therefore
\begin{equation}
\kappa_r=\frac{\omega_r^2 M^2}{L R_{\rm eff}}
=\frac{2\omega_r^3 M^2}{\pi Z_r R_0}.
\label{eq:kappa_mutual}
\end{equation}
The lumped-element model is validated by finite-element electromagnetic simulation results as shown in Fig.~\ref{fig:concept}(f), in which we
extract $\kappa_r=\mathrm{Re}(Y_{\rm in})/C_r$ at $\omega_r$,
where $Y_{\rm in}$ is the simulated load admittance referred to
the dissipator port with the mutual-coupling transformation included,
and $C_r$ is the corresponding equivalent
capacitance~\cite{neeley2008transformed}.
For the design parameters $\omega_r/2\pi=3.7\,\mathrm{GHz}$ and $M=140\,\mathrm{pH}$, at the matched design point $R_0=50\,\Omega$, the analytical model and simulations give $\kappa_r/2\pi=10.0$ and $9.7~\mathrm{MHz}$, respectively.
The decay rate, however, can be affected by the impedance mismatching between the transmission line and the terminal resistance. 
The lumped model predicts that $\kappa_r/2\pi$ remains within $8.3$--$12.5~\mathrm{MHz}$ for termination resistances within $\pm20\%$ of the nominal $50~\Omega$ value [Fig.~\ref{fig:concept}(f)].
For a finite reset line (approximately $19~\mathrm{mm}$ in our device), termination mismatch makes $\kappa_r$ depend on the dissipator coupling position. See the Supplemental Material~\cite{supplemental_material} for details.

\begin{figure}[!t]
\centering
\includegraphics[width=\columnwidth]{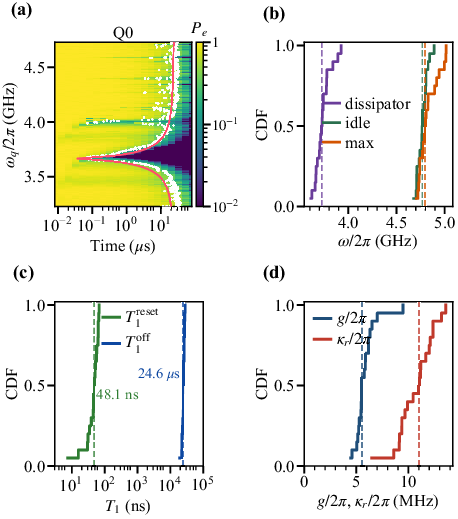}
\caption{Characterization of resonator-mediated reset.
(a)~Excited-state population $P_e$ versus hold time and qubit frequency $\omega_q/2\pi$ for Q0. White markers show $T_1(\omega_q)$ from exponential fits. The red curve is a fit of the total-decay model, yielding $g/2\pi=5.9~\mathrm{MHz}$ and $\kappa_r/2\pi=8.6~\mathrm{MHz}$.
(b)~CDFs of the dissipator frequencies $\omega_r/2\pi$ (purple), idle frequencies $\omega_{\rm idle}/2\pi$ (green), and maximum qubit frequencies $\omega_{\rm max}/2\pi$ (orange), with respective medians of $3.724$, $4.768$, and $4.799~\mathrm{GHz}$.
(c)~CDFs of reset decay times $T_1^{\rm reset}$ (green) and off-resonant lifetimes $T_1^{\rm off}$ (blue).
(d)~CDFs of the fitted coupling strengths $g/2\pi$ (blue) and dissipator decay rates $\kappa_r/2\pi$ (red), with respective medians of $5.5$ and $11.0~\mathrm{MHz}$. Dashed lines in (b)--(d) mark the medians.}
\label{fig:eg_reset}
\end{figure}

\begin{figure}[!t]
\centering
\includegraphics[width=\columnwidth]{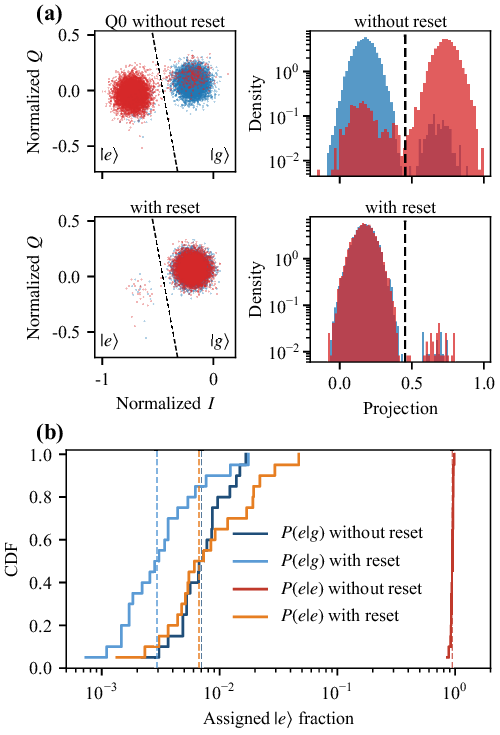}
\caption{Single-shot characterization of qubit reset.
(a)~Single-shot $I$--$Q$ distributions (left) and projected histograms (right) for Q0 without reset (top) and with a $400~\mathrm{ns}$ reset pulse (bottom). Blue (red) denotes preparation in $|g\rangle$ ($|e\rangle$). Reset reduces $P(e|g)$ from $1.39\%$ to $0.18\%$ and $P(e|e)$ from $95.78\%$ to $0.37\%$. Dashed lines indicate the fixed discrimination threshold determined without reset.
(b)~CDFs of $P(e|g)$ (dark/light blue) and $P(e|e)$ (red/orange), with colors listed in the order without/with reset. Reset reduces the median $P(e|g)$ from $0.70\%$ to $0.29\%$ and the median $P(e|e)$ from $94.74\%$ to $0.67\%$. Dashed lines mark the medians.}
\label{fig:reset_singleshot}
\end{figure}

To characterize the reset performance across multiple qubits, we measure the relaxation dynamics of 20 operational qubits on a 72-qubit test processor, whose qubits, couplers, and resonators follow a design similar to that reported previously~\cite{chen2025efficient}. To separate the crosstalk effect, the qubits are divided into four groups (see the Supplemental Material~\cite{supplemental_material}), with simultaneous state preparation and readout performed within each group. Each qubit is initialized in $|e\rangle$ at its idle frequency $\omega_{\rm idle}$ and subjected to a flat-top flux pulse with cosine-shaped rising and falling edges totaling $2~\mathrm{ns}$. The pulse tunes the qubit to a variable frequency $\omega_q$, holds it for a duration $t$, and returns it to $\omega_{\rm idle}$. 
At each hold frequency, we extract the relaxation time $T_1(\omega_q)$ by fitting the resulting excited-state population $P_e(\omega_q,t)$ as a function of $t$ to an exponential decay [white markers in Fig.~\ref{fig:eg_reset}(a)].
Figure~\ref{fig:eg_reset}(a) shows the results for a representative qubit, which exhibits strongly enhanced relaxation near the dissipator frequency $\omega_r$. 
In contrast, at large detunings (typically $>500~\mathrm{MHz}$ in our experiment), the relaxation time becomes nearly frequency independent, indicating effective suppression of dissipator-induced Purcell decay.

We define the minimum fitted relaxation time as $T_1^{\rm reset}$ and use the corresponding qubit frequency to identify $\omega_r$. To determine the qubit--dissipator coupling strength $g$ and dissipator decay rate $\kappa_r$, we model the resonator-induced qubit decay rate as~\cite{heinsoo2018rapid,zhou2021rapid,supplemental_material}
\begin{equation}
\kappa_q(\Delta)
=\frac{1}{2}\left[
\kappa_r-\mathrm{Re}\sqrt{(\kappa_r+2i\Delta)^2-16g^2}
\right],
\label{eq:kappa_q}
\end{equation}
where $\Delta=\omega_q-\omega_r$ denotes the qubit--dissipator detuning. Assuming a frequency-independent background relaxation rate $\Gamma_{\rm bg}=1/T_1^{\rm off}$ [Fig.~\ref{fig:eg_reset}(c)], the total relaxation rate is given by $1/T_1(\omega_q)=\kappa_q(\Delta)+\Gamma_{\rm bg}$. Fitting this expression to the measured $T_1(\omega_q)$, with $\Gamma_{\rm bg}$ and $\omega_r$ fixed, yields $g$ and $\kappa_r$ [red curve in Fig.~\ref{fig:eg_reset}(a)]. See the Supplemental Material~\cite{supplemental_material} for details. For the analyzed qubits, the extracted parameters satisfy $|g|>\kappa_r/4$, placing the system in the strong-coupling regime, where Eq.~\eqref{eq:kappa_q} predicts a resonant decay envelope with rate $\kappa_q(0)=\kappa_r/2$.

Figures~\ref{fig:eg_reset}(b)--(d) summarize the measured and extracted parameters across the 20 qubits as cumulative distribution functions (CDFs). The median relaxation time is reduced from $T_1^{\rm off}=24.6~\mu\mathrm{s}$ off resonance to $T_1^{\rm reset}=48.1~\mathrm{ns}$ during reset [Fig.~\ref{fig:eg_reset}(c)]. The dissipator frequencies have a median of $\omega_r/2\pi=3.724~\mathrm{GHz}$, while the median idle and maximum qubit frequencies are $\omega_{\rm idle}/2\pi=4.768~\mathrm{GHz}$ and $\omega_{\rm max}/2\pi=4.799~\mathrm{GHz}$, respectively [Fig.~\ref{fig:eg_reset}(b)]. The idle frequencies are chosen near the maximum frequencies. The extracted qubit--dissipator coupling strengths and dissipator decay rates have medians of $g/2\pi=5.5~\mathrm{MHz}$ and $\kappa_r/2\pi=11.0~\mathrm{MHz}$, respectively [Fig.~\ref{fig:eg_reset}(d)].

The median extracted coupling strength is lower than the design value of $8.5~\mathrm{MHz}$, which corresponds to a qubit--dissipator coupling capacitance of $C_{qr}=1.2~\mathrm{fF}$.
This difference is potentially due to fabrication variations in the small coupling capacitance or systematic uncertainties in parameter extraction. In particular, qubit dephasing~\cite{gao2026nonlocal,supplemental_material} and residual flux-pulse distortions despite infinite impulse response (IIR) correction~\cite{rol2020timedomain}, neither of which is included in the coupled-mode model, may influence the extracted parameters and reset dynamics. Such pulse distortions may also affect the short-time population dynamics and bias the fitted $T_1^{\rm reset}$, potentially accounting for part of the discrepancy between measured and model-predicted reset times.

The same model [Eq.~\eqref{eq:kappa_q}] suggests a route to improving frequency selectivity.
For the design $\kappa_r/2\pi=10.0~\mathrm{MHz}$, reducing $g/2\pi$ from $8.5$ to $2.5~\mathrm{MHz}$ would reach critical damping, $4g=\kappa_r$, the smallest coupling that retains the maximum resonant rate.
Within this model, the detuning range over which the dissipator-induced lifetime $1/\kappa_q$ falls below $1~\mathrm{ms}$ would narrow from $|\Delta|/2\pi\lesssim2.1~\mathrm{GHz}$ to $|\Delta|/2\pi\lesssim0.63~\mathrm{GHz}$.
This would expand the frequency range available for qubit operation while keeping dissipator-induced relaxation below the specified threshold.
A second resonator between the qubit and dissipator could further improve frequency selectivity~\cite{heinsoo2018rapid}.

\begin{figure}[!t]
\centering
\includegraphics[width=\columnwidth]{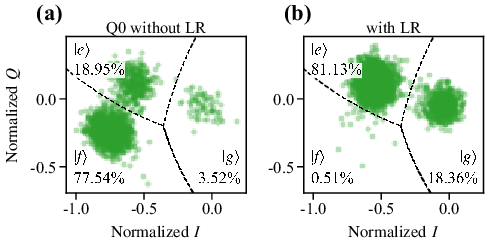}
\caption{Leakage removal (LR) from $|f\rangle$ on Q0. Single-shot $I$--$Q$ distributions after $|f\rangle$ preparation (a)~without and (b)~with a $400~\mathrm{ns}$ leakage-removal pulse. The measurements use 2048 shots per prepared state without leakage removal and 4096 with leakage removal. Dashed curves show the decision boundaries of a Gaussian mixture model trained without leakage removal and reused in both panels. Percentages give the assignment probabilities for the indicated states.}
\label{fig:fe_reset}
\end{figure}

We assess reset performance across the measured qubits using single-shot readout with and without a $400~\mathrm{ns}$ reset pulse, acquiring 8192 shots per prepared state ($|g\rangle$ or $|e\rangle$) per condition [Fig.~\ref{fig:reset_singleshot}(a)].
During reset, each qubit is tuned into resonance with its dissipator to activate $|e\rangle\!\to\!|g\rangle$ decay.
For each qubit, the discrimination threshold is determined without reset and held fixed for both conditions.
We denote by $P(i|j)$ the probability of assigning $|i\rangle$ after preparation in $|j\rangle$.

Figure~\ref{fig:reset_singleshot}(b) summarizes the assignment probabilities across the measured qubits.
The median $P(e|g)$ decreases from $0.70\%$ ($0.52$--$0.89\%$) without reset to $0.29\%$ ($0.17$--$0.46\%$) with reset, while the median $P(e|e)$ decreases from $94.74\%$ ($92.88$--$95.57\%$) to $0.67\%$ ($0.47$--$1.75\%$).
Parentheses give the 25th--75th percentile ranges across qubits.
The higher median and broader interquartile range following $|e\rangle$ preparation could partly arise from excitation temporarily stored in spurious modes, such as two-level-system (TLS) defects, and subsequently returned to the qubit during flux tuning~\cite{kim2025fast}.
We define the reset efficiency as the fractional reduction in excited-state assignment probability, $\eta=1-P_{\rm reset}(e|e)/P_{\rm no~reset}(e|e)$. Evaluating $\eta$ for each qubit gives a median of $99.28\%$, confirming the effectiveness of our scheme leveraging on-chip dissipation for passive reset.

To estimate an effective temperature, we assume that the assigned fractions approximate thermal populations and use $p_e/p_g=\exp[-\hbar\omega/(k_{\mathrm{B}}T_{\rm eff})]$.
Using the median values, $P_{\rm no~reset}(e|g)=0.70\%$ at $\omega_{\rm idle}/2\pi=4.768~\mathrm{GHz}$ corresponds to $T_{\rm eff}=46.2~\mathrm{mK}$, whereas $P_{\rm reset}(e|g)=0.29\%$ evaluated at the reset frequency $\omega_r/2\pi=3.724~\mathrm{GHz}$ gives $T_{\rm eff}=30.6~\mathrm{mK}$.
This interpretation assumes equilibration during reset ($\approx8T_1^{\rm reset}$), negligible readout bias, and negligible population changes before readout.
The lower inferred temperature could reflect reduced exposure to external noise, while its excess above the refrigerator base temperature ($\approx10~\mathrm{mK}$) may indicate imperfect thermalization or residual radiation.
Improved thermal anchoring and shielding could further reduce excitation~\cite{malevannaya2025engineering}.

We extend the protocol to leakage removal from $|f\rangle$ on Q0 (Fig.~\ref{fig:fe_reset}).
A $400~\mathrm{ns}$ flux pulse brings the $|f\rangle\leftrightarrow|e\rangle$ transition into resonance with the dissipator, enhancing $|f\rangle\!\to\!|e\rangle$ decay.
For three-state single-shot discrimination, we use a spherical Gaussian mixture model (GMM) trained without leakage removal and apply the same classifier to both datasets.
Following $|f\rangle$ preparation, the pulse reduces $P(f|f)$ from $77.54\%$ to $0.51\%$.
The dominant post-pulse assignment to $|e\rangle$ is consistent with resonator-mediated $|f\rangle\!\to\!|e\rangle$ population transfer.
The accompanying increase in $P(g|f)$, from $3.52\%$ without the pulse to $18.36\%$ after the pulse, may reflect additional relaxation during the pulse sequence.
Possible contributions include residual Purcell-mediated $|e\rangle\!\to\!|g\rangle$ decay and other dissipative channels encountered during flux tuning, such as the relaxation feature near $\omega_q/2\pi\approx4~\mathrm{GHz}$ in Fig.~\ref{fig:eg_reset}(a), which may be associated with TLS defects.

In summary, we demonstrate fast superconducting qubit reset using dedicated dissipative resonators coupled to shared on-chip resistive loads, without requiring additional reset-specific cryogenic ports or control requirements. We achieve a median reset time constant of $T_1^{\rm reset}=48.1~\mathrm{ns}$. A $400~\mathrm{ns}$ reset pulse yields median residual excited-state populations of $0.29\%$ and $0.67\%$ for qubits initialized in $|g\rangle$ and $|e\rangle$, respectively. We further demonstrate efficient leakage removal on Q0, reducing the residual $|f\rangle$-state population to $0.51\%$ with a $400~\mathrm{ns}$ pulse. Further improvements could be achieved through cryogenic characterization and optimization of the impedance-matched terminations, together with joint optimization of $g$ and $\kappa_r$ to balance reset speed and frequency selectivity. Beyond initialization and leakage removal, on-chip resistive loads provide a scalable resource for engineered dissipation~\cite{harrington2022engineered}, with potential applications in autonomous quantum error correction~\cite{kapit2016hardware,gertler2021protecting} and dissipative preparation of quantum-correlated many-body states~\cite{mi2024stable}.

\begin{acknowledgments}
This work was supported by the Beijing Natural
Science Foundation (Grants No. 4262019 and No. JQ25014), the National Natural Science Foundation of China (Grants No. 12322413 and No. 92476206), and the Beijing Academy of Quantum Information Sciences (Grant No. Y26NG262803).
\end{acknowledgments}


\bibliography{apssamp}

\clearpage
\begin{bibunit}[apsrev4-2]
\makeatletter
\let\supp@natanchorstart\hyper@natanchorstart
\let\supp@natlinkstart\hyper@natlinkstart
\let\supp@natlinkbreak\hyper@natlinkbreak
\renewcommand{\hyper@natanchorstart}[1]{\supp@natanchorstart{supp.#1}}
\renewcommand{\hyper@natlinkstart}[1]{\supp@natlinkstart{supp.#1}}
\renewcommand{\hyper@natlinkbreak}[2]{\supp@natlinkbreak{#1}{supp.#2}}
\makeatother
\onecolumngrid
\begin{center}
{\Large\bfseries Supplemental Material: Implementing rapid reset of superconducting qubits with on-chip dissipation\par}
\vspace{0.8em}
{\normalsize Feiyu Li$^{1,2,3,*}$, Yang Liu$^{1,*}$, Zhen Yang$^{1,*}$, Zhenxing Liu$^1$, Jiayu Ding$^1$, Yang Gao$^{1,2,3}$, Xuehui Liang$^1$, Weijie Sun$^1$, Yanjun Ma$^1$, Chuhong Yang$^1$, Guangming Xue$^{1,\dagger}$, He Wang$^{1,\dagger}$, and Fei Yan$^{1,\dagger}$\par}
\vspace{0.5em}
{\small\itshape
$^1$Beijing Key Laboratory of Fault-Tolerant Quantum Computing, Beijing Academy of Quantum Information Sciences, Beijing 100193, China\par
$^2$Beijing National Laboratory for Condensed Matter Physics, Institute of Physics, Chinese Academy of Sciences, Beijing 100190, China\par
$^3$University of Chinese Academy of Sciences, Beijing 101408, China\par}
{\small $^*$These authors contributed equally to this work.\par}
{\small $^\dagger$Corresponding authors: xuegm@baqis.ac.cn, wanghe@baqis.ac.cn, and yanfei@baqis.ac.cn.\par}
\end{center}
\twocolumngrid
\setcounter{section}{0}
\renewcommand{\theHsection}{supp.\arabic{section}}
\renewcommand{\theHequation}{supp.\arabic{equation}}
\renewcommand{\theHfigure}{supp.\arabic{figure}}
\renewcommand{\theHtable}{supp.\arabic{table}}

\renewcommand{\theequation}{S\arabic{equation}}
\renewcommand{\thefigure}{S\arabic{figure}}
\renewcommand{\thetable}{S\arabic{table}}
\setcounter{equation}{0}
\setcounter{figure}{0}
\setcounter{table}{0}

This Supplemental Material describes the circuit models, reset dynamics, and measurement protocol, and examines pure dephasing. We use angular-frequency units for Hamiltonians ($\hbar=1$).

\section{Circuit model of the reset line}
\label{sec:reset-line-circuit}
\subsection{Input impedances of the two line segments}
\label{sec:line-segments}
Consider a uniform, lossless transmission line with characteristic impedance $Z_t$, phase velocity $v_p=c/\sqrt{\epsilon_{\rm eff}}$, and total length $\ell$. Both ends are terminated by $R_0$. The dissipator couples locally to the line at position $x$, leaving segments of length $x$ and $\ell-x$. We consider mutual-inductive and capacitive couplers below. We use phasors with time dependence $e^{i\omega t}$ and evaluate the weak-loading response at the dissipator frequency $\omega_r$.

For a segment of length $d$ terminated by $R_0$, the input impedance at phase constant $\beta=\omega_r/v_p$ is
\begin{equation}
Z_{\rm in}(d)=Z_t\frac{R_0\cos(\beta d)+iZ_t\sin(\beta d)}{Z_t\cos(\beta d)+iR_0\sin(\beta d)}.
\label{eq:finite-zin}
\end{equation}

\subsection{Mutual-inductive loading and dissipator decay}
In the localized series-coupling representation used here, the induced electromotive force is inserted in the reset-line conductor. Its return path passes through the two line arms and their grounded terminations. The impedance seen across this series source is therefore
\begin{equation}
Z_\Sigma(x)=Z_{\rm in}(x)+Z_{\rm in}(\ell-x).
\label{eq:finite-zsum}
\end{equation}

Let $I_r$ be the peak phasor of the dissipator current at its inductive coupling point. The induced line current $I_t=-i\omega_r M I_r/Z_\Sigma$ gives a reflected voltage $i\omega_r M I_t=Z_{\rm refl}I_r$, with
\begin{equation}
Z_{\rm refl}(x)=\frac{(\omega_r M)^2}{Z_\Sigma(x)}.
\end{equation}
The real part supplies dissipation; the imaginary part produces a reactive frequency shift, neglected when evaluating the loading at the specified $\omega_r$. The time-averaged dissipated power is
\begin{equation}
P_{\rm loss}=\frac12|I_r|^2\operatorname{Re}Z_{\rm refl}.
\end{equation}
For the $\lambda/4$ dissipator, the equivalent inductance referred to its short-end current is $L_r=\pi Z_r/(4\omega_r)$. Its total stored energy is $U=L_r|I_r|^2/2$ at resonance. Thus the energy-decay rate is
\begin{equation}
\kappa_r(x,R_0)=\frac{(\omega_r M)^2}{L_r}\operatorname{Re}\!\left[\frac{1}{Z_{\rm in}(x)+Z_{\rm in}(\ell-x)}\right].
\label{eq:finite-kappa}
\end{equation}
Here $\kappa_r$ is an energy decay rate.

\subsubsection{Matched termination and zero-length limit}
For matched terminations ($R_0=Z_t$), Eq.~\eqref{eq:finite-zin} gives $Z_{\rm in}(d)=R_0$ at any segment length. In the zero-electrical-length limit ($\beta\ell\to0$), $Z_{\rm in}(d)$ also approaches $R_0$, irrespective of matching. Both cases give $Z_\Sigma=2R_0$ and hence
\begin{equation}
\kappa_r=\frac{\omega_r^2M^2}{2L_rR_0}.
\label{eq:matched-Lr}
\end{equation}
Substituting the quarter-wave inductance $L_r=\pi Z_r/(4\omega_r)$ referred to the short-end current gives
\begin{equation}
\kappa_r=\frac{2\omega_r^3M^2}{\pi Z_rR_0},
\label{eq:kappa-maintext-limit}
\end{equation}
which is Eq.~(1) of the main text.

\subsubsection{Impedance mismatch and coupling-position dependence}
\begin{figure}[!htbp]
\centering
\includegraphics[width=\columnwidth]{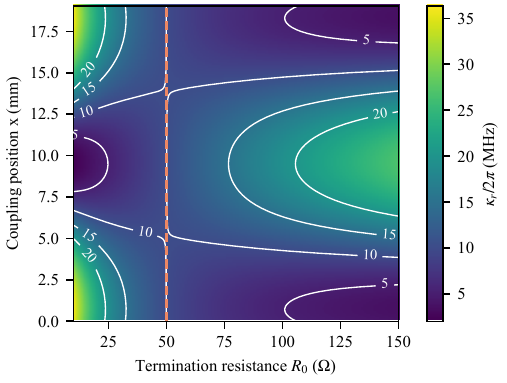}
\caption{Dissipator decay rate $\kappa_r/2\pi$ as a function of the resistance $R_0$ of each termination and the coupling position $x$ along a $19~\mathrm{mm}$ reset line. The dashed line marks impedance matching ($R_0=Z_t$), where $\kappa_r/2\pi=9.98~\mathrm{MHz}$ independent of $x$. The calculation uses $\omega_r/2\pi=3.7~\mathrm{GHz}$, $M=140~\mathrm{pH}$, $Z_r=Z_t=50~\Omega$, and $\epsilon_{\rm eff}=5.3$. White contours are labeled in MHz. The model assumes a lossless line and a single localized coupler.}
\label{fig:finite-line-map}
\end{figure}

Away from matching, the termination reflection coefficient $\Gamma_L=(R_0-Z_t)/(R_0+Z_t)$ acquires different round-trip phases in the two arms. Equivalently,
\begin{equation}
Z_{\rm in}(d)=Z_t\frac{1+\Gamma_L e^{-2i\beta d}}{1-\Gamma_L e^{-2i\beta d}}.
\end{equation}
The transformed load makes $\kappa_r$ position dependent, with $\kappa_r(x)=\kappa_r(\ell-x)$ for equal terminations. Figure~\ref{fig:finite-line-map} shows this dependence at fixed $\ell$.

\subsection{Capacitive loading and dissipator decay}
A capacitor $C_c$ couples the dissipator voltage antinode to the reset line. For matched terminations ($R_0=Z_t$), the two line segments each present $R_0$ and are in parallel. The dissipator therefore sees the series impedance $(i\omega_r C_c)^{-1}+R_0/2$, independent of coupling position and line length. Taking the real part of its admittance and dividing by $C_r$ gives
\begin{equation}
\kappa_r^{(C)}=
\frac{\omega_r^2 C_c^2 R_0}{2C_r[1+(\omega_r C_c R_0/2)^2]}
\simeq \frac{2\omega_r^3 Z_r C_c^2 R_0}{\pi},
\end{equation}
where the last step uses $\omega_r C_c R_0/2\ll1$ and $C_r=\pi/(4\omega_r Z_r)$.

\section{Dissipator decay rate from $S$-parameters}
\label{sec:port-admittance}
Let $Y_{\rm in}$ be the simulated load admittance referred to the dissipator voltage antinode, including the coupling transformation, and $C_r$ the equivalent capacitance at that plane. The energy-decay rate is~\cite{neeley2008transformed}
\begin{equation}
\kappa_r=\frac{\operatorname{Re}Y_{\rm in}}{C_r}.
\end{equation}
With $S_{11}$ referenced to the dissipator characteristic impedance $Z_r$,
\begin{equation}
Y_{\rm in}=\frac{1}{Z_r}\frac{1-S_{11}}{1+S_{11}}.
\end{equation}
Using $C_r=\pi/(4\omega_r Z_r)$ for a quarter-wave resonator at its voltage antinode gives
\begin{equation}
\kappa_r=\frac{4\omega_r}{\pi}\frac{1-|S_{11}|^2}{|1+S_{11}|^2}.
\label{eq:kappa-port-general}
\end{equation}
In the weak-coupling limit at the chosen reference plane, $S_{11}\simeq+1$. Equation~\eqref{eq:kappa-port-general} then simplifies to
\begin{equation}
\kappa_r\simeq\frac{\omega_r}{\pi}\left(1-|S_{11}|^2\right).
\label{eq:kappa-port-weak}
\end{equation}
With the same reference-plane and energy-normalization conventions, this method applies to both inductive and capacitive coupling.

\section{Parameter dependence of the dissipator decay rate}
\subsection{Frequency dependence}
\begin{figure}[!htbp]
\centering
\includegraphics[width=\columnwidth]{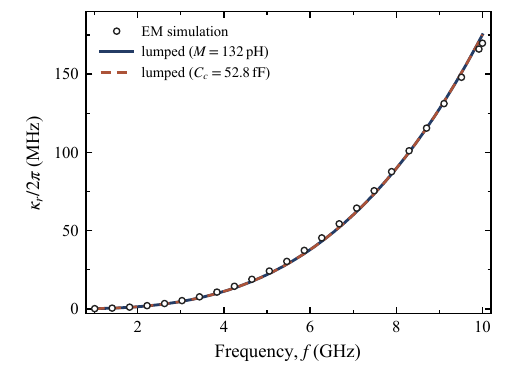}
\caption{Frequency-dependent dissipator loading: electromagnetic simulations (open markers) and matched lumped models with mutual-inductive coupling ($M=132~\mathrm{pH}$, solid) and capacitive coupling ($C_c=52.8~\mathrm{fF}$, dashed). Both use $Z_r=R_0=50~\Omega$.}
\label{fig:kappa-frequency}
\end{figure}

For matched terminations and weak loading, both mutual-inductive and capacitive coupling yield $\kappa_r\propto\omega_r^3$. Their decay rates are identical when $C_c=M/(Z_rR_0)$. Figure~\ref{fig:kappa-frequency} compares this dependence with electromagnetic simulations over 1--10 GHz. The fitted values are $M=132~\mathrm{pH}$ and $C_c=52.8~\mathrm{fF}$.

\subsection{Dependence on termination resistance}
\begin{figure}[!htbp]
\centering
\includegraphics[width=\columnwidth]{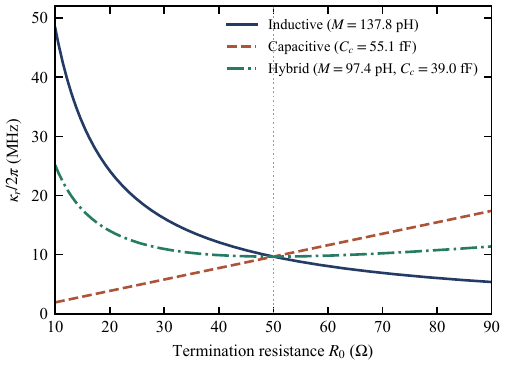}
\caption{Calculated dissipator decay rates versus the resistance of each termination. The three models share a reference rate $\kappa_*/2\pi=9.67$ MHz at $R_0=Z_t=50~\Omega$. The legend gives the coupler parameters for $\omega_r/2\pi=3.7$ GHz and $Z_r=50~\Omega$. The balanced hybrid curve is the sum of equal inductive and capacitive contributions at matching.}
\label{fig:hybrid-resistance}
\end{figure}

Consider the zero-electrical-length model near the matched design point $R_0=Z_t$. In the weak-coupling limit, the inductive contribution scales as $1/R_0$ and the capacitive contribution scales as $R_0$. To illustrate resistance tolerance, assume the two contributions add in rate and choose their strengths so that they contribute equally at matching. If $\kappa_*$ is the total rate at matching and $r=R_0/Z_t$, the hybrid rate is
\begin{equation}
\kappa_{\rm hyb}(r)=\frac{\kappa_*}{2}(r+r^{-1}).
\end{equation}
At $r=1$, the linear sensitivity to $R_0$ vanishes. Figure~\ref{fig:hybrid-resistance} compares the resulting quadratic variation near matching with the purely inductive and capacitive cases.

\section{Qubit decay}
\label{sec:hybridized-decay}
\subsection{Decay rate and population dynamics}
In the single-excitation sector, we neglect intrinsic qubit relaxation and pure dephasing and take the dissipator bath to be in vacuum. The homogeneous Langevin equations in the qubit frame are $\dot b=-iga$ and $\dot a=(i\Delta-\kappa_r/2)a-igb$, where $\Delta=\omega_q-\omega_r$. The slower amplitude pole gives the population-envelope rate used in Eq.~(2) of the main text~\cite{heinsoo2018rapid,zhou2021rapid},
\begin{equation}
\kappa_q(\Delta)=\frac12\left[\kappa_r-\operatorname{Re}\sqrt{(\kappa_r+2i\Delta)^2-16g^2}\right].
\label{eq:derived-kappaq}
\end{equation}
On resonance, $\kappa_q(0)=\kappa_r/2$ for $4|g|\ge\kappa_r$. Far off resonance, $\kappa_q\simeq\kappa_r(g/\Delta)^2$.

For an initial state $|e,0\rangle$, the same equations give
\begin{equation}
P_e(t)=e^{-\kappa_r t/2}\left|\cosh\frac{Dt}{4}
+\frac{\kappa_r-2i\Delta}{D}\sinh\frac{Dt}{4}\right|^2,
\label{eq:exact-reset-population}
\end{equation}
where $D=\sqrt{(\kappa_r-2i\Delta)^2-16g^2}$.

\subsection{Effect of qubit dephasing}
\label{sec:qubit-dephasing}
The coherent population dynamics in Eq.~\eqref{eq:exact-reset-population} exhibit damped qubit--dissipator oscillations, whereas the measured $P_e(t)$ traces are approximately described by the exponential fit in Eq.~\eqref{eq:pe-row}. Pure dephasing suppresses this oscillatory contrast. Dephasing may therefore contribute to the approximately exponential decay observed experimentally, although its rate during reset has not been independently measured.

We integrate the resonant Jaynes--Cummings master equation with $H=g(\sigma_+a+\sigma_-a^\dagger)$ and dissipative terms~\cite{blais2021circuit}
\begin{equation}
\mathcal L[\rho]=\kappa_r\mathcal D[a]\rho+\frac{\Gamma_\phi}{2}\mathcal D[\sigma_z]\rho,
\label{eq:lindblad-deph}
\end{equation}
where $\mathcal D[L]\rho=L\rho L^\dagger-\{L^\dagger L,\rho\}/2$. The pure-dephasing rate $\Gamma_\phi$ is quoted in units of $10^6~\mathrm{s}^{-1}$.

\begin{figure}[!htbp]
\centering
\includegraphics[width=\columnwidth]{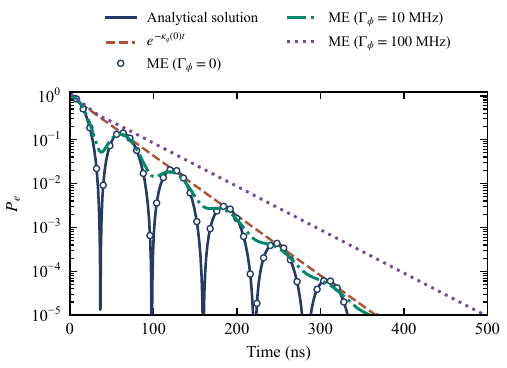}
\caption{Resonant qubit excitation for $g/2\pi=8.5$ MHz and $\kappa_r/2\pi=10$ MHz. The solid curve is the analytical population, the open circles are the numerical master-equation solution, and the dashed curve is $e^{-\kappa_q(0)t}$ with $\kappa_q(0)=\kappa_r/2$. Dash-dotted and dotted curves show master-equation results for $\Gamma_\phi=10$ and $100$ MHz, respectively. The bath is in vacuum and intrinsic qubit relaxation is omitted.}
\label{fig:analytical-reset}
\end{figure}

In Fig.~\ref{fig:analytical-reset}, the analytical and Lindblad solutions agree at $\Gamma_\phi=0$ and show coherent exchange oscillations, with $e^{-\kappa_q(0)t}$ setting the decay scale. Pure dephasing reduces the oscillation contrast and, at $\Gamma_\phi=100$ MHz, slows qubit decay~\cite{gao2026nonlocal}.

\section{Chip layout and measurement protocol}
\begin{figure}[!htbp]
\centering
\includegraphics[width=\columnwidth,height=0.50\textheight,keepaspectratio]{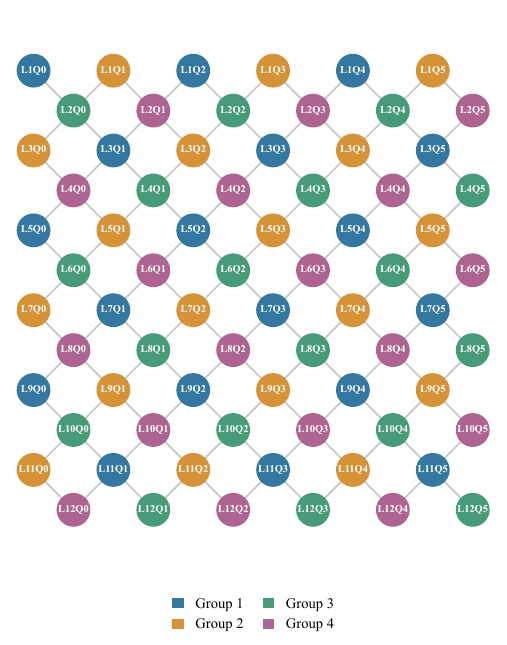}
\caption{Schematic layout of the 72-qubit processor. Circles are qubits labeled by reset line and position within that line; thin gray lines indicate nearest-neighbor connections. Four colors identify the four geometric groups, each containing 18 sites.}
\label{fig:chip-layout}
\end{figure}

The processor contains 72 qubits arranged in the staggered layout shown in Fig.~\ref{fig:chip-layout}. Qubits are labeled L$x$Q$y$, where $x=1,\ldots,12$ identifies the reset line and $y=0,\ldots,5$ identifies the qubit on that line. Each circle represents one qubit; thin gray lines indicate nearest-neighbor connections. The grid is partitioned into four nonadjacent groups. Within each row, even and odd columns belong to different groups. For each group, the selected column parity alternates between successive rows of the same row parity. Each group contains 18 sites.

Preparation, reset, and readout are performed in groups to reduce crosstalk. The reported distributions combine measurements of the 20 analyzed qubits from seven lines.

\begin{figure}[!htbp]
\centering
\includegraphics[width=\columnwidth]{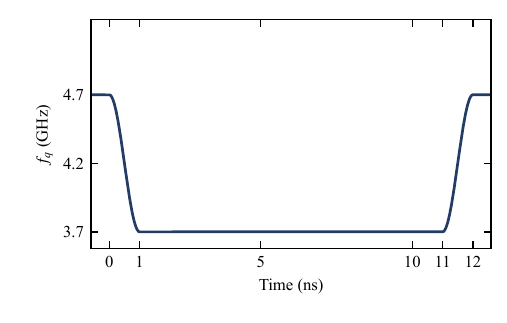}
\caption{Illustrative cosine-edge reset waveform with a 1 ns falling edge, a 10 ns plateau, and a 1 ns rising edge. The qubit frequency is tuned from an idle value of 4.7 GHz to a reset frequency of 3.7 GHz and back. }
\label{fig:reset-waveform}
\end{figure}

The reset waveform consists of a cosine falling edge, a constant-frequency plateau, and a cosine rising edge, as shown in Fig.~\ref{fig:reset-waveform}. Each edge lasts $\tau=1$ ns. Taking $s=0$ at the start of the falling edge, the qubit frequency is
\begin{equation}
 f_q(s)=\begin{cases}
 f_r+\dfrac{A_f}{2}[1+\cos(\pi s/\tau)],&0\le s<\tau,\\
 f_r,&\tau\le s<\tau+T_p,\\
 f_r+\dfrac{A_f}{2}[1-\cos(\pi u/\tau)],&0\le u<\tau,
 \end{cases}
\label{eq:reset-waveform}
\end{equation}
where $A_f=f_{\rm idle}-f_r$ and $u=s-\tau-T_p$ on the rising edge. Outside the pulse, $f_q=f_{\rm idle}$. The illustrative waveform in Fig.~\ref{fig:reset-waveform} uses $T_p=10$ ns so that both edges and the plateau are visible on a common time axis. Sweeping the plateau frequency and duration produces the reset maps $P_e(\omega_q,t)$. The illustrative trajectory tunes the qubit from 4.7 GHz to 3.7 GHz and back.

\section{Extraction of experimental parameters}
\label{sec:reset-map-analysis}
\begin{figure*}[!tp]
\centering
\includegraphics[width=\textwidth,height=0.87\textheight,keepaspectratio]{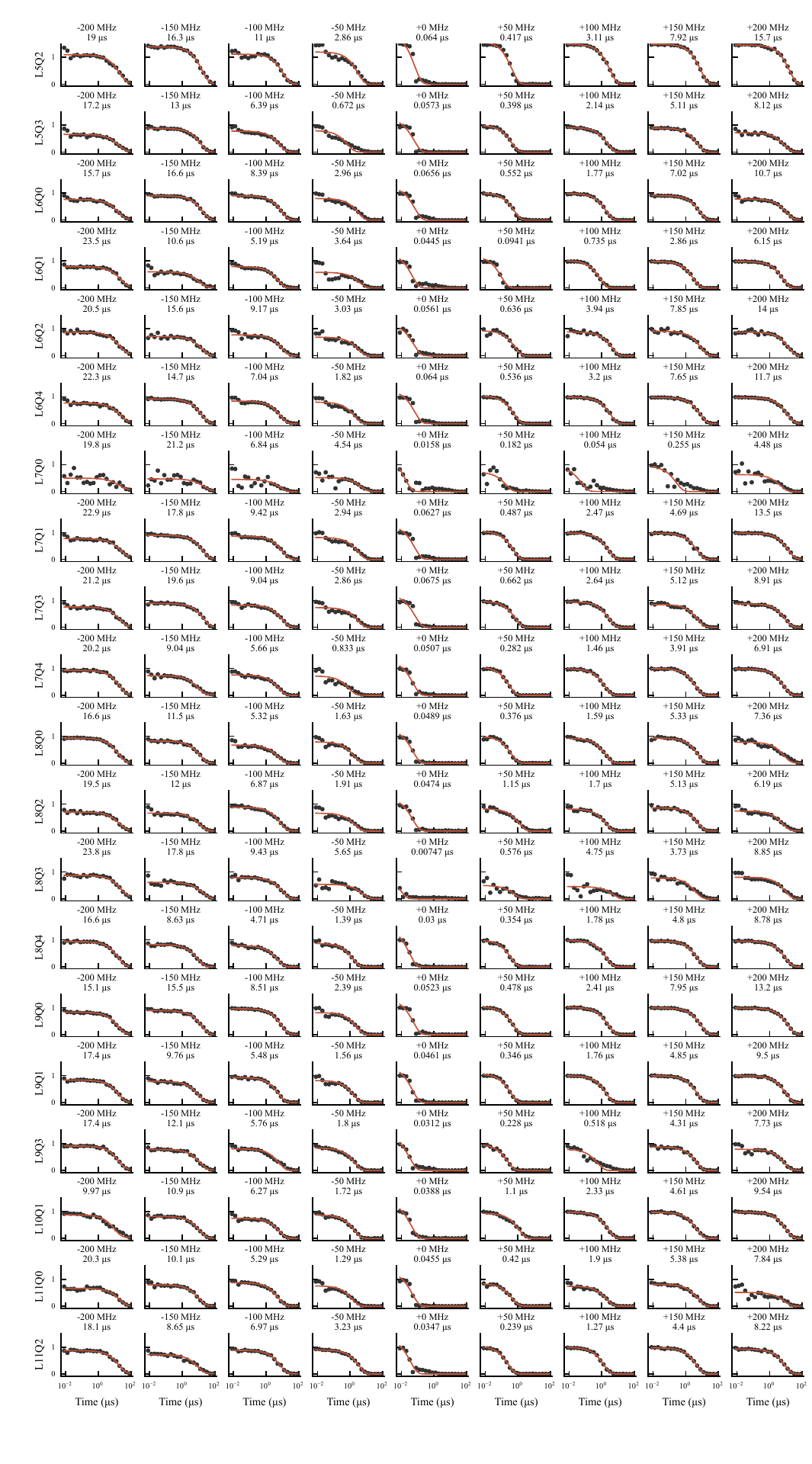}
\caption{Exponential fits to $P_e(t)$ for all 20 qubits at nine measured frequencies nearest the target detunings. Each row is one qubit. Points are measurements and curves are fits to Eq.~\eqref{eq:pe-row} with fixed offset. Panel titles give detuning and fitted lifetime.}
\label{fig:t1-traces}
\end{figure*}

At each measured hold frequency, we fit
\begin{equation}
P_e(t)=c+A\exp(-t/T_1),
\label{eq:pe-row}
\end{equation}
with offset $c$ fixed. 
The use of an exponential fit rather than Eq.~\eqref{eq:exact-reset-population} is discussed in Sec.~\ref{sec:qubit-dephasing}. 
Figure~\ref{fig:t1-traces} shows exponential fits within $\pm200$ MHz of reset in 50 MHz steps.

\begin{figure*}[!tp]
\centering
\includegraphics[width=\textwidth,height=0.68\textheight,keepaspectratio]{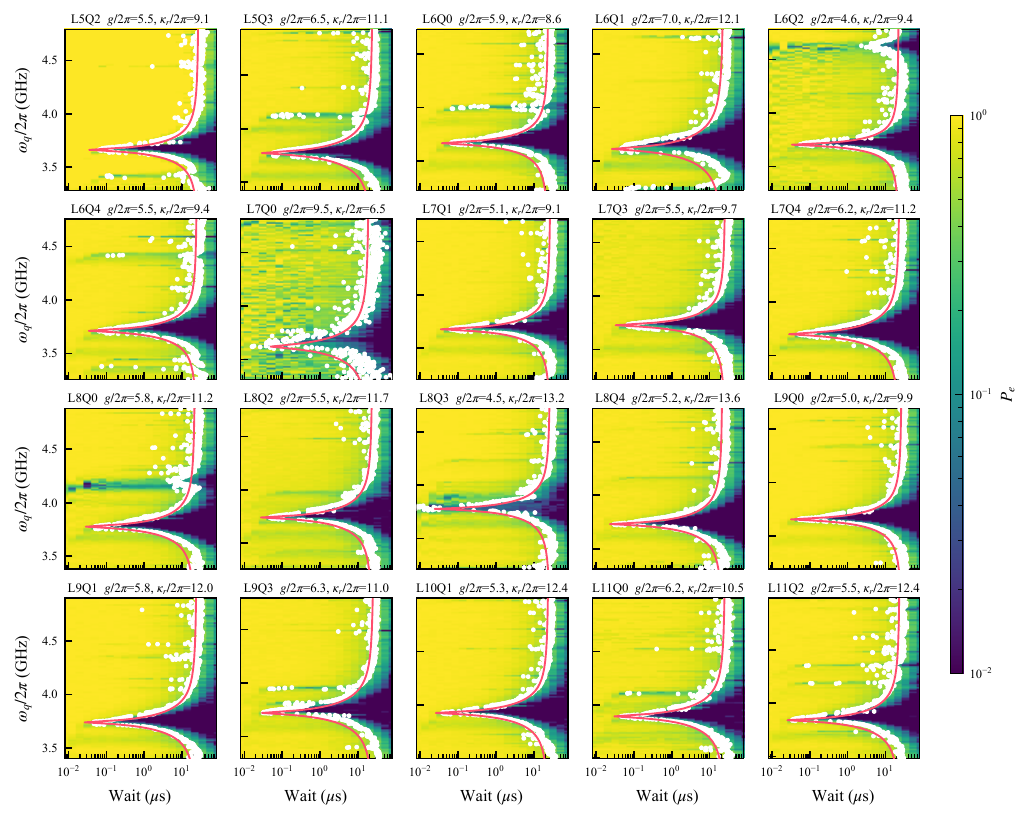}
\caption{Reset maps for the main-text ensemble. White markers are fixed-offset exponential lifetimes; red curves use the main-text fit with both $g$ and $\kappa_r$ free.}
\label{fig:pe2d-t1points}
\end{figure*}

We extract $g$ and $\kappa_r$ from these $T_1$ markers using the method adopted in the main text. With $\Delta=\omega_q-\omega_r$, the total-decay model is
\begin{equation}
T_1^{-1}(\omega_q)=\kappa_q(\Delta)+\Gamma_{\rm bg},\qquad
\Gamma_{\rm bg}=1/T_1^{\rm off},
\end{equation}
where $\kappa_q(\Delta)$ is given by Eq.~\eqref{eq:derived-kappaq}. The reset frequency is fixed at the frequency of the shortest fitted lifetime, and the background rate is fixed. Both $g$ and $\kappa_r$ are fitted in $\log T_1$. Figure~\ref{fig:pe2d-t1points} shows the population maps, exponential $T_1$ markers, and fitted lifetime curves for all 20 qubits. The median values are $g/2\pi=5.5~\mathrm{MHz}$ and $\kappa_r/2\pi=11.0~\mathrm{MHz}$, with a measured reset-lifetime median of $48.1~\mathrm{ns}$.

\begin{figure}[!htbp]
\centering
\includegraphics[width=\columnwidth]{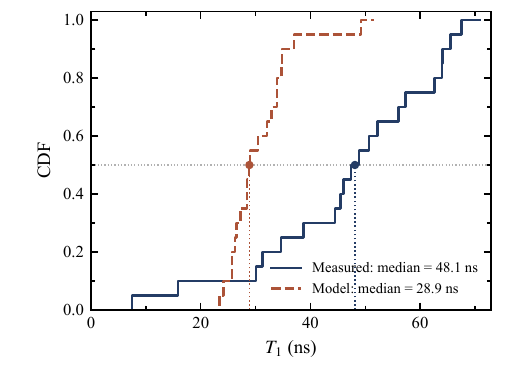}
\caption{CDFs of the measured reset lifetimes and the resonant lifetimes predicted from the fitted $g$ and $\kappa_r$ for the same 20 qubits. Legends give the medians, $48.1$ and $28.9$ ns, with dots marking their positions.}
\label{fig:t1-meas-vs-model}
\end{figure}

The measured reset lifetime is the minimum exponential lifetime in the reset-frequency window. The model-predicted resonant lifetime is
\begin{equation}
T_{1,\mathrm{model}}^{\rm reset}=\left[\kappa_q(0)+\Gamma_{\rm bg}\right]^{-1},
\end{equation}
evaluated using each qubit's fitted $g$ and $\kappa_r$. Figure~\ref{fig:t1-meas-vs-model} compares the two CDFs for the same 20 qubits. The measured median reset lifetime is $48.1~\mathrm{ns}$, compared with the model prediction of $28.9~\mathrm{ns}$. As illustrated in Sec.~\ref{sec:qubit-dephasing}, pure dephasing can suppress coherent exchange oscillations and slow qubit relaxation during reset. Residual flux-pulse distortions after infinite impulse response correction~\cite{rol2020timedomain} may also alter the short-time population dynamics and bias the exponential lifetime fits. These effects, which are omitted from the coupled-mode fit, may contribute to the discrepancy.

\ifdefined\StandaloneSupplement
\renewcommand{\bibsection}{\section*{Supplemental References}}
\bibliographystyle{apsrev4-2}
\bibliography{apssamp}
\else
\begingroup
\renewcommand{\bibsection}{\section*{Supplemental References}}
\let\supplementbiblabel\label
\renewcommand{\label}[1]{\supplementbiblabel{supp-bib:#1}}
\putbib[apssamp]
\endgroup
\fi

\end{bibunit}

\end{document}